\documentclass{article}

\usepackage{PRIMEarxiv}

\usepackage[utf8]{inputenc} 
\usepackage[T1]{fontenc}    
\usepackage{hyperref}       
\usepackage{url}            
\usepackage{booktabs}       
\usepackage{amsfonts}       
\usepackage{nicefrac}       
\usepackage{microtype}      
\usepackage{lipsum}
\usepackage{graphicx}
\graphicspath{{media/}}     

\usepackage{float}

\usepackage{tabularx}
\usepackage[table]{xcolor}
\usepackage{geometry}
\usepackage{listings}
\usepackage{xcolor} 

\lstdefinelanguage{SQL}{
  morekeywords={SELECT, FROM, WHERE, GROUP, BY, ORDER, AS, AVG, TO_CHAR},
  sensitive=false,
  morecomment=[l]{--},
  morestring=[b]',
}

\title{Technical Implementation of Tippy: Multi-Agent Architecture and System Design for Drug Discovery Laboratory Automation
}

\author{
\parbox{\textwidth}{
\centering
Yao Fehlis, Charles Crain, Aidan Jensen, Michael Watson, James Juhasz, Paul Mandel, Betty Liu, Shawn Mahon, Daren Wilson, Nick Lynch-Jonely, Ben Leedom, David Fuller\\
Artificial, Inc.
}
}

\begin{document}
\maketitle

\begin{abstract}
Building on the conceptual framework presented in our previous work\cite{fehlis2025accelerating} on agentic AI for pharmaceutical research, this paper provides a comprehensive technical analysis of Tippy's multi-agent system implementation for drug discovery laboratory automation. We present a distributed microservices architecture featuring five specialized agents (Supervisor, Molecule, Lab, Analysis, and Report) that coordinate through OpenAI Agents SDK orchestration and access laboratory tools via the Model Context Protocol (MCP)\cite{anthropic_mcp_2024}. The system architecture encompasses agent-specific tool integration, asynchronous communication patterns, and comprehensive configuration management through Git-based tracking. Our production deployment strategy utilizes Kubernetes container orchestration with Helm charts, Docker containerization, and CI/CD pipelines for automated testing and deployment. The implementation integrates vector databases for RAG functionality and employs an Envoy reverse proxy for secure external access. This work demonstrates how specialized AI agents can effectively coordinate complex laboratory workflows while maintaining security, scalability, reliability, and integration with existing laboratory infrastructure through standardized protocols.
\end{abstract}


\section{Introduction}
The transition from conceptual agentic AI frameworks to production-ready laboratory automation systems requires addressing numerous technical challenges related to system architecture, protocol design, and integration with existing laboratory infrastructure. While our previous work established the scientific rationale and conceptual framework for multi-agent laboratory automation, this paper focuses on the engineering decisions, implementation details, and technical characteristics that enable practical deployment of such systems.

Modern laboratory automation demands real-time coordination between heterogeneous systems, robust error handling, and seamless integration with both automated instrumentation and human workflows. Traditional laboratory management systems provide data storage and basic workflow tracking but lack the intelligence and coordination capabilities required for autonomous operation across complex multi-phase workflows like the Design-Make-Test-Analyze (DMTA) cycle\cite{plowright2012hypothesis}.

Our implementation addresses these limitations through a distributed multi-agent architecture that combines specialized domain expertise with flexible coordination mechanisms. Each agent operates autonomously within its domain while participating in coordinated workflows that span multiple disciplines and laboratory systems.

\section{Multi-Agent Architecture and Specialized Capabilities}
\label{sec:method}

\subsection{Agent Specialization and Domain Expertise}

\begin{figure}[H]
  \centering
  \includegraphics[width=1.0\textwidth]{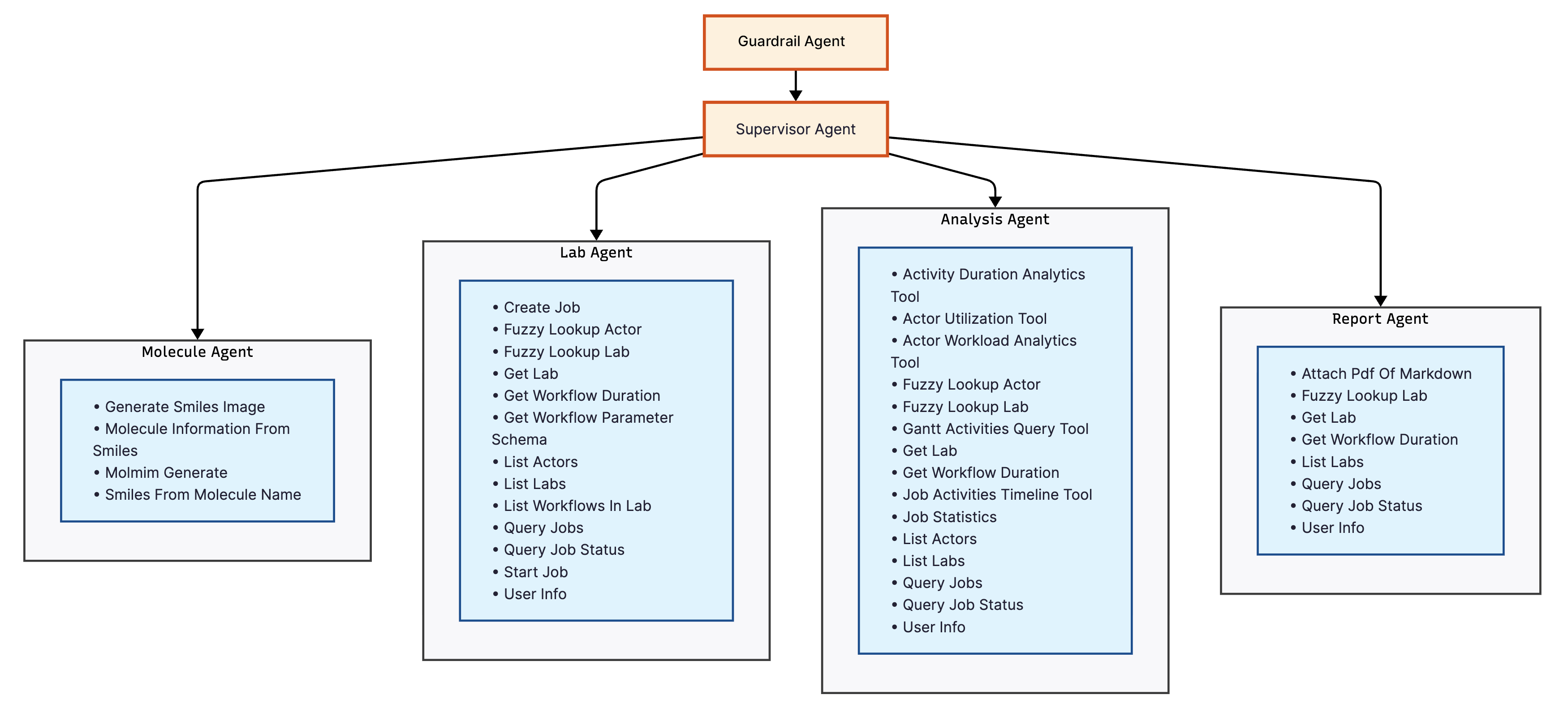} 
  \caption{Multi-agent architecture showing the Supervisor Agent coordinating with four specialized agents - Molecule, Lab, Analysis, and Report agents - along with the Safety Guardrail Agent for laboratory safety validation.}
  \label{fig:fig1}
\end{figure}

The system employs a supervisor-agent pattern with five specialized agents, each equipped with domain-specific tools accessed through the Model Context Protocol (MCP). As illustrated in Figure~\ref{fig:fig1}, the architecture demonstrates clear separation of concerns and efficient tool utilization across the laboratory automation workflow.

\subsection{Supervisor Agent Coordination}
The Supervisor Agent serves as the central coordinator, leveraging OpenAI Agents SDK's handoff mechanisms to orchestrate workflows across four specialized agents: Molecule, Lab, Analysis, and Report agents. The system employs sophisticated context sharing and dynamic agent routing that enable complex decision-making across multiple experimental phases.

\subsection{Molecule Agent: Computational Chemistry Integration}
The Molecule Agent focuses specifically on molecular and chemical compound operations through 4 specialized MCP tools:
\begin{itemize}
  \item \textbf{Generate SMILES Image}: Creates molecular structure visualizations
  \item \textbf{Molecule Information from SMILES}: Retrieves detailed chemical properties from SMILES notation
  \item \textbf{Molmim Generate}: Leverages the MolMIM model\cite{reidenbach2022improving} for property-guided molecular generation
  \item \textbf{SMILES from Molecule Name}: Converts chemical names to standardized SMILES notation
\end{itemize}

The agent exemplifies the modular design approach, implementing specialized interfaces for molecular representation handling (SMILES formats), property prediction using the MolMIM model running on NVIDIA GPUs, and synthesis planning through retrosynthetic analysis tools. Function calling mechanisms enable dynamic tool selection based on molecular complexity and project requirements.

Molecular representation handling supports multiple formats including SMILES and proprietary structure formats used by computational chemistry software. The agent can convert between formats automatically and maintains consistency across different tools and databases while leveraging the MolMIM model for molecular generation.

Property-guided molecular generation capabilities utilize the MolMIM model to generate molecules that maximize user-specified scoring functions. The agent can generate molecular designs against multiple criteria including drug-like properties, toxicity profiles, and activity targets, with GPU acceleration enabling rapid processing of complex molecular generation tasks.

\subsection{Lab Agent: Workflow Orchestration and Instrumentation Control}

The Lab Agent contains the most extensive tool set with 13 MCP tools, including comprehensive laboratory management capabilities:
\begin{itemize}
  \item \textbf{Attach PDF of Markdown}: Document attachment functionality
  \item \textbf{Create Job}: Initiates new laboratory workflows and experiments
  \item \textbf{Fuzzy Lookup Actor}: Identifies the most likely workflow-associated actor (equipment or humans) from a non-exact human-specified input string
  \item \textbf{Fuzzy Lookup Lab}: Locates specific laboratory spaces and configurations
  \item \textbf{Get Lab}: Retrieves detailed laboratory information and status
  \item \textbf{Get Workflow Duration}: Tracks timing and performance metrics
  \item \textbf{Get Workflow Parameter Schema}: Accesses workflow configuration templates
  \item \textbf{List Actors}: Lists workflow-associated actors (equipment or humans)
  \item \textbf{List Labs}: Catalogs available laboratory facilities
  \item \textbf{List Workflows in Lab}: Shows available experimental procedures
  \item \textbf{Query Jobs}: Searches and filters laboratory job records
  \item \textbf{Query Job Status}: Monitors real-time job progress and completion
  \item \textbf{Start Job}: Executes laboratory workflows and instrument control
  \item \textbf{User Info}: Retrieves operator credentials and permissions
\end{itemize}

The Lab Agent functions as the primary interface to the Artificial platform\cite{fehlis2025accelerating1}, managing HPLC analysis workflows, synthesis procedures, and laboratory job execution. It can create and start jobs, query job status and results, manage workflow parameters, and coordinate laboratory resources. The agent understands JSON schema requirements for job parameters but communicates with scientists in accessible language, handling technical details behind the scenes.

Protocol interpretation capabilities allow the agent to analyze experimental procedures and suggest modifications to protocols that reduce reagent consumption, minimize instrument time, or improve data quality.

\subsection{Analysis Agent: Data Processing and Pattern Recognition}
The Analysis Agent is equipped with 10 comprehensive MCP tools for data analysis and laboratory insights:
\begin{itemize}
  \item \textbf{Attach PDF of Markdown}: Creates analytical reports and documentation
  \item \textbf{Fuzzy Lookup Actor}: Identifies the most likely workflow-associated actor (equipment or humans) from a non-exact human-specified input string
  \item \textbf{Fuzzy Lookup Lab}: Locates laboratories for performance analysis
  \item \textbf{Get Lab}: Retrieves laboratory configuration data for analysis
  \item \textbf{Get Workflow Duration}: Analyzes timing patterns and efficiency metrics
  \item \textbf{List Actors}: Lists workflow-associated actors (equipment or humans)
  \item \textbf{List Labs}: Catalogs laboratories for comparative studies
  \item \textbf{Query Jobs}: Performs complex searches across experimental datasets
  \item \textbf{Query Job Status}: Queries the current status of laboratory jobs
  \item \textbf{User Info}: Accesses operator data for performance correlation
\end{itemize}

The Analysis Agent serves as a specialized data analyst, processing job performance data and extracting statistical insights from laboratory workflows. It has access to advanced analytics tools including activity duration analysis, actor workload monitoring, and job timeline tracking\cite{fehlis2025uncovering}. It uses retention time data from HPLC analysis to guide molecular design decisions, as retention time correlates with key drug properties, feeding this information back to the Molecule Agent for design guidance.

Statistical analysis capabilities include hypothesis testing, regression analysis, and multivariate statistics appropriate for pharmaceutical research applications. The agent can automatically select appropriate statistical methods based on data characteristics and research objectives.

\subsection{Report Agent: Documentation and Communication}
The Report Agent specializes in documentation and reporting with 8 MCP tools focused on data retrieval and formatting:
\begin{itemize}
  \item \textbf{Attach PDF of Markdown}: Converts markdown reports to PDF format
  \item \textbf{Fuzzy Lookup Lab}: Locates laboratories for reporting context
  \item \textbf{Get Lab}: Retrieves laboratory details for comprehensive reporting
  \item \textbf{Get Workflow Duration}: Includes timing metrics in performance reports
  \item \textbf{List Labs}: Catalogs laboratories for multi-site reporting
  \item \textbf{Query Jobs}: Gathers experimental data for report generation
  \item \textbf{Query Job Status}: Queries the current status of laboratory jobs
  \item \textbf{User Info}: Includes operator information in experimental documentation
\end{itemize}

The Report Agent acts as a documentation specialist, generating summary reports and detailed scientific documentation from experimental data. It can convert findings into PDF formats and attach reports directly to job results in the Artificial Lab suite, ensuring that insights from experiments are properly captured and shared with research teams.

\subsection{Safety Guardrail Agent: Content Filtering and Safety Validation}

The Safety Guardrail Agent implements OpenAI's content filtering and moderation systems to ensure safe and appropriate interactions within the laboratory automation environment. This agent operates through built-in content moderation rather than external MCP tools, providing input and output filtering to detect and prevent potentially harmful content, inappropriate requests, or safety policy violations. The system includes topical guardrails, prompt injection detection, and content moderation filters that scan for problematic categories while maintaining compliance with laboratory safety standards.

\section{System Architecture and Technical Implementation}
\begin{figure}[H]
  \centering
  \includegraphics[width=1.0\textwidth]{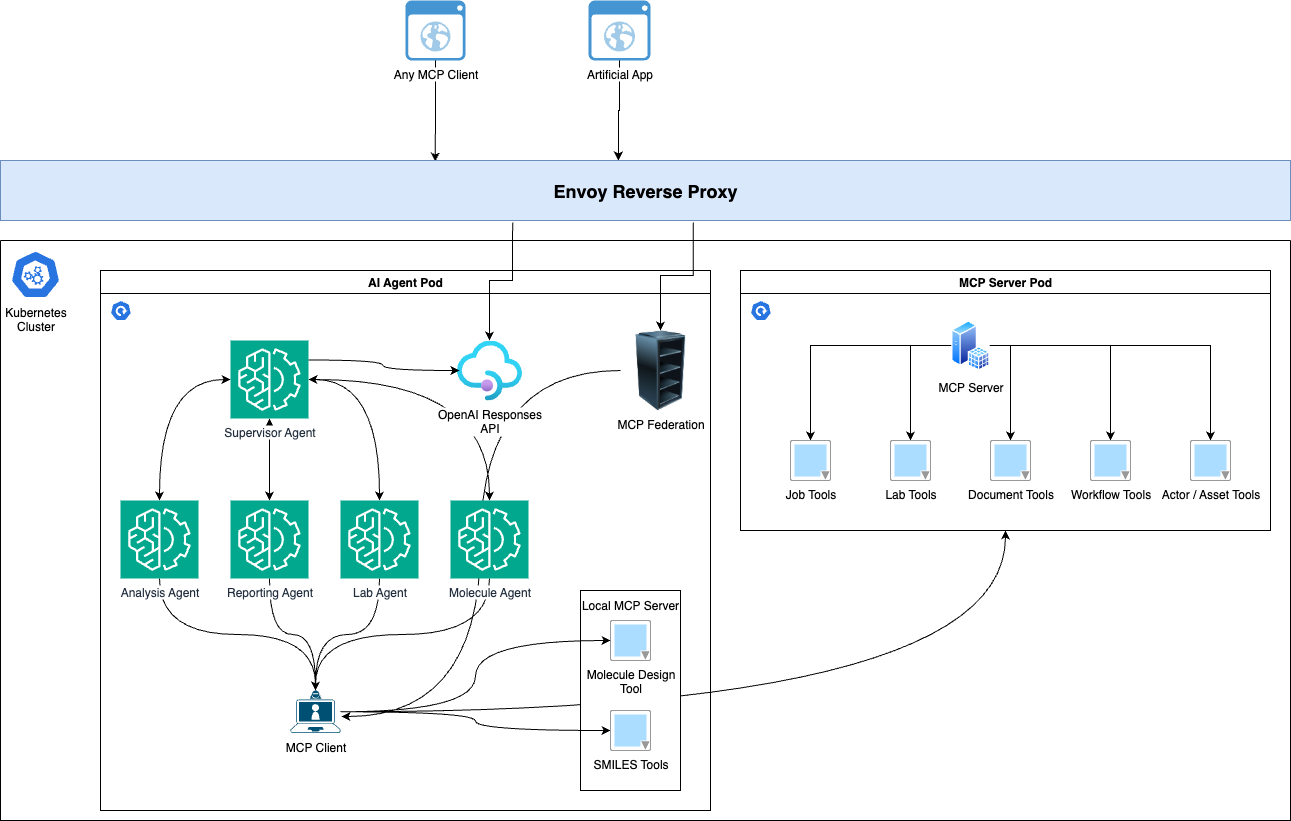} 
  \caption{System architecture diagram showing the Kubernetes-deployed infrastructure with AI Agent Pod, MCP Server Pod, external client access via Envoy proxy, and MCP tool integration - note that the Safety Guardrail Agent is not included in this system architecture diagram as it operates through built-in validation rather than external infrastructure.}
  \label{fig:fig2}
\end{figure}

The system implements a production-grade multi-agent architecture using OpenAI Agents SDK as the core orchestration framework, designed to provide scalable and resilient scientific workflow automation\cite{kandogan2025orchestrating}. As illustrated in Figure~\ref{fig:fig2} the platform employs a distributed microservices design with specialized components for agent coordination, tool integration, and user interaction, all deployed on a Kubernetes cluster.

External users access the system through Any MCP Client or the Artificial App, both communicating with the platform via an Envoy reverse proxy that routes requests to appropriate services within the cluster. The core agentic intelligence resides in the AI Agent Pod, containing the Supervisor Agent that orchestrates task delegation across specialized agents (Molecule, Lab, Analysis, and Report agents). These agents coordinate using the OpenAI Responses API to access foundational models for reasoning and language understanding, while also interacting with the MCP Federation for distributed coordination with other MCP Servers.

The architecture leverages modern LLM agent patterns including sophisticated prompt engineering, function calling mechanisms, and multi-agent collaboration protocols. The system implements retrieval-augmented generation (RAG) through integrated vector databases for long-term memory and context management, enabling agents to maintain persistent memory across experimental campaigns and learn from historical data patterns.

\subsection{Model Context Protocol Integration}

The Model Context Protocol (MCP) serves as the primary mechanism for integrating AI agents with laboratory tools, analytical instruments, and data systems. The system architecture implements this through a dedicated MCP Server Pod that houses the MCP Server, providing access to five categories of specialized tools:
\begin{itemize}
  \item \textbf{Job Tools}: Workflow management and task scheduling
  \item \textbf{Lab Tools}: Laboratory equipment control and monitoring
  \item \textbf{Document Tools}: File management and documentation
  \item \textbf{Workflow Tools}: Process automation and coordination
  \item \textbf{Actor/Asset Tools}: Resource management and allocation
\end{itemize}

The MCP Client serves as the user interface for interacting with both AI agents and server-side tools, completing the data and action flow loop between automated reasoning and experimental execution. The client can send structured requests to the MCP Server or connect to Local MCP Servers that host domain-specific tools like Molecule Design Tool and SMILES Tools. Each MCP tool exposes a well-defined interface including input/output specifications, error conditions, and operational constraints, enabling agents to reason about tool capabilities and compose complex workflows.

\subsection{Agent Coordination and Handoff}

Multi-agent coordination employs OpenAI Agents SDK's native handoff and context sharing features to ensure consistent information flow and intelligent task routing. The system implements context window management strategies while maintaining relevant historical context for decision-making across agent transitions.

The asynchronous communication patterns enable agents to continue processing while waiting for responses from other agents or tools, maximizing system throughput by avoiding blocking operations. This approach is particularly important for laboratory workflows where agents coordinate across different time scales - from rapid molecular calculations to longer-running experimental procedures.

\section{Agent Management and Observability}

\subsection{Configuration Management}
The system implements comprehensive configuration management for agent deployments. All agent configuration changes and prompt engineering experiments are automatically logged with detailed metadata including hyperparameters and configuration parameters.

Data management employs Git-based tracking to manage experimental datasets, molecular libraries, and analysis results across different research campaigns. This enables reproducible agent behavior and facilitates tracking of configuration changes across different deployments.

The system maintains detailed lineage tracking for all agent configurations, including prompt templates, tool configurations, and agent coordination policies. Git-based versioning with Git tags ensures that all changes to agent behavior can be traced back to specific code commits and configuration changes, enabling precise deployment tracking and rollback capabilities.

\subsection{Runtime Monitoring and Observability}
The system implements comprehensive runtime monitoring using OpenAI Tracing for observability across all agent interactions and tool usage patterns. This provides detailed insights into agent behavior, tool utilization, and workflow execution patterns across the laboratory automation environment.

Context window management strategies maintain relevant historical context for decision-making across agent transitions. The observability framework enables real-time tracking of agent performance and interaction patterns throughout complex laboratory workflows.

\section{Production Deployment and Infrastructure}

\subsection{Kubernetes Deployment with Helm}
Helm charts provide declarative configuration management for Kubernetes deployments, enabling consistent and repeatable deployments across development, staging, and production environments. The charts support configurable scaling, resource allocation, and integration settings.

The deployment architecture includes separate pods for each agent type with configurable resource limits and scaling policies. Horizontal Pod Autoscaling (HPA)\cite{nguyen2020horizontal} automatically adjusts agent counts based on CPU utilization and custom metrics such as message queue depth.

Configuration management through ConfigMaps and Secrets enables environment-specific customization without rebuilding container images. Database connections, API keys, and tool configurations are externalized and managed through Kubernetes native mechanisms.

Rolling updates and blue-green deployment strategies ensure zero-downtime deployments and rapid rollback capabilities. The system can update individual agent types without disrupting overall workflow execution.

\subsection{CI/CD Pipeline Integration}
The production deployment leverages GitHub Actions for comprehensive CI/CD pipelines that integrate application deployment workflows. The pipeline implements automated testing of tool integration and end-to-end workflow validation across multiple cloud environments.

Docker-based application packaging automatically builds and publishes container images to cloud registries, with security scanning and vulnerability assessment integrated into the build process. Kubernetes job orchestration manages distributed deployment tasks, leveraging GPU clusters for agent inference workloads.

The system implements blue-green deployment strategies with automated rollback capabilities, ensuring zero-downtime updates for production agent deployments. Container orchestration provides reliable deployment and scaling of agent services across the Kubernetes cluster.

\section{Conclusion and Future Directions}
The technical implementation of multi-agent systems for laboratory automation represents a significant advancement in pharmaceutical research infrastructure. The combination of specialized agent capabilities, OpenAI Agents SDK orchestration, and robust MCP integration enables autonomous coordination of complex laboratory workflows while maintaining the flexibility and oversight required for scientific research.

Our production-grade implementation demonstrates substantial improvements in workflow efficiency, resource utilization, and decision quality compared to traditional approaches. The system successfully addresses key technical challenges including scalability, reliability, and integration with existing laboratory infrastructure through containerized deployment and comprehensive agent configuration management practices.

Future enhancements will focus on sophisticated human-in-the-loop capabilities that enable seamless collaboration between AI agents and laboratory researchers. This integration will provide mechanisms for human oversight, intervention, and guidance during critical decision points, ensuring that complex scientific decisions benefit from both AI efficiency and human expertise. Interactive approval workflows will enable agents to request human confirmation for high-stakes experimental decisions, safety-critical operations, or novel experimental approaches outside their training data.

\bibliographystyle{unsrt}  
\bibliography{references}

\end{document}